\documentclass[twocolumn,showpacs,prb,10pt]{revtex4}
\usepackage{graphicx}

\begin{document}

\title {Finite temperature transport in disordered Heisenberg chains}

\author{A. Karahalios, A. Metavitsiadis, X. Zotos$^1$,}
\affiliation{$^1$ Department of Physics, University of Crete and
Foundation for Research and Technology-Hellas, P.O. Box 2208, 71003
Heraklion, Greece}
\author{A. Gorczyca$^{2,3}$, and P. Prelov\v sek$^{2,4}$}
\affiliation{$^2$J.\ Stefan Institute,
SI-1000 Ljubljana, Slovenia}
\affiliation{$^3$ Department of Theoretical Physics, Institute of Physics, 
University of Silesia, 40-007 Katowice, Poland}
\affiliation{$^4$ Faculty of Mathematics
and Physics, University of Ljubljana, SI-1000 Ljubljana, Slovenia}
\date{\today}

\begin{abstract}
Using numerical diagonalization techniques, we explore the effect of
local and bond disorder on the finite temperature spin and thermal
conductivities of the one dimensional anisotropic spin-1/2 Heisenberg
model. High-temperature results for local disorder show 
that the dc conductivies are finite, 
apart from the uncorrelated - XY case - where dc transport vanishes. 
Moreover, at strong disorder, we find finite
dc conductivities at all temperatures $T$, except $T=0$. 
The low frequency conductivities are characterized by a 
nonanalytic cusp shape. Similar behavior is found for bond disorder.
\end{abstract}

\maketitle

The effect of correlations on localized states in disordered
system\cite{g4} is a long standing problem\cite{pwa} that is
attracting renewed theoretical and experimental
interest\cite{nat,mirlin,alt,huse2,dhar}.  While it is clear that in a
one-dimensional (1D) noninteracting disordered system all states are
localized, the introduction of correlations might lead to different
possible scenaria.  First, at zero temperature, $T=0$, numerical
results for fermions with repulsive interaction in a disordered system
reveal that localization persists in spite of
correlations\cite{eckern,doty}, although some types of interactions
might destroy the localized states leading the system to a normal
diffusive state or even one with diverging low frequency
conductivity\cite{huse1}.  Even if the system remains localized at
$T=0$, an arbitrary low temperature could delocalize it or a finite
critical temperature\cite{mirlin,alt} might be needed to drive it to a
normal state at high temperatures. There are also indications that in
the presence of large disorder even at high $T$ many body states can
appear effectively localized\cite{huse2,zpp}.

To explore this issue, the 1D (in general) anisotropic spin-$1/2$
Heisenberg chain is a minimal model that allows to investigate the
interplay of disorder and correlations on transport.  In the XY limit,
mapped by the Jordan-Wigner transformation to a system of
noninteracting spinless fermions\cite{xy}, it is expected to have all
single-particle states localized under any amount of local or bond
disorder, consistent with the Anderson localization
phenomenon\cite{g4}.  The spin (in fermionic representation equivalent
to charge) conductivity as well as the thermal d.c. conductivities are
expected to vanish at all $T$, while a.c. conductivities are finite but 
non-trivial\cite{albers}. 

Without disorder, the anisotropic XXZ model is a strongly correlated
spin system, with nearest neighbor interaction in the fermion picture.
It is integrable using the Bethe ansatz method for any value of the
anisotropy and it is known to show ideal spin/charge (in the
easy-plane case) and thermal (for any anisotropy) conductivities at
all temperatures\cite{znp}.

Besides the theoretical interest of this model quasi-1D magnetic
compounds have recently been synthesized which are described
exceedingly well by the 1D isotropic spin-1/2 Heisenberg model and
show unusually high thermal conductivity due to a magnetic transport
mode contribution\cite{hess}.  Relevant to this work, experiments are
underway to study the effect of disorder by non-magnetic as well as
magnetic impurities.

In this work, we will use state of the art numerical diagonalization
techniques - the exact diagonalization (ED), the finite-temperature
Lanczos method (FTLM) \cite{ftlm} and the microcanonical Lanczos
method (MCLM)\cite{mclm} - to see what they can offer on this issue of
disorder and correlations.  While we will study the spin and thermal
conductivity of the spin-$1/2$ anisotropic Heisenberg model, the spin
conductivity maps directly to that of the charge conductivity of a
spinless fermion model.

We first consider the 1D anisotropic spin-$1/2$ Heisenberg model in
the presence of a random local magnetic field,

\begin{equation}
H=\sum_l J (S^x_lS^x_{l+1}+S^y_lS^y_{l+1}+\Delta S^z_lS^z_{l+1}) + \sum_l
b_l S^z_l, \label{sham}
\end{equation}
where $S^{\alpha}, \alpha=x,y,z$ are spin-1/2 operators, $J$ is the
magnetic exchange coupling, $\Delta$ the anisotropy parameter and
$-W/2 < b_l < +W/2$ random local fields from a uniform distribution.
We assume periodic boundary conditions, $\hbar=\kappa_B=1$ and take
$J$ as the unit of energy.

Our analysis will be based on standard linear response theory.
The spin conductivity is given by
\begin{eqnarray}
\sigma(\omega)&=&\frac{1}{\omega L}\Im \int_0^{+\infty} dt e^{izt} < [j(t),j)]>,
\nonumber\\
j&=&\sum_l j_{l,l+1}= J\sum_l (S^x_lS^y_{l+1}-S^y_l S^x_{l+1}), \label{sig}
\end{eqnarray}
$j$ representing the spin current. The corresponding thermal
conductivity is given by,
\begin{equation} 
\kappa(\omega)=\frac{\beta}{\omega L}\Im \int_0^{+\infty} dt e^{izt} 
< [j^{\epsilon}(t),j^{\epsilon})]>, \label{kappa}
\end{equation}
where $\beta=1/T$. The energy current $j^\epsilon$ in an
inhomogeneous system can be defined via the dipole operator\cite{maha}
\begin{equation}
P^\epsilon= \sum_l r_l h_l, \qquad j^\epsilon=i\sum_{lm} r_l[h_m,h_l],  
\label{je}
\end{equation}
whereby $h_l$ are local energy operators and $r_l$ the correponding
coordinates.  Taking into account that locations of local field
energies are on sites and exchange energies on bonds, respectively,
one arrives at
\begin{eqnarray}
j^{\epsilon}&=&J^2\sum_l j^{\epsilon}_l+ \frac{b_l+b_{l+1}}{2} j_{l,l+1},
\nonumber\\
j^{\epsilon}_l &=&
(S^x_{l+1}S^z_lS^y_{l-1}-S^y_{l+1}S^z_l S^x_{l-1})\nonumber\\
&+&\Delta (x\rightarrow y,z; y\rightarrow z,x; z\rightarrow x,y).  
\end{eqnarray}

For vanishing random fields, $W=0$, the energy current $j^\epsilon$
commutes with the Hamiltonian for all values of the anisotropy
$\Delta$ and thus the system is an ideal thermal conductor\cite{znp}
at all $T$. It has also been shown that the uniform, $W=0$, system
exhibits ballistic spin transport in the XY regime, $\Delta <1$, at
all temperatures although $j$ does not commute with $H$ \cite{z}.

Let us start the analysis with the high-$T$ limit, $T \rightarrow
\infty$, which exhibits nontrivial $\sigma(\omega)$ and
$\kappa(\omega)$, being in fact quite generic for all $T>0$.  The
difference between spin and thermal transport can be realized already
from the spin, $M^s_n=\int \omega^n \sigma(\omega) d\omega =\pi\beta
m_n^s$ and corresponding energy, $M_n^\epsilon=\pi\beta^2m_n^\epsilon$ 
frequency moments.
Moments can be evaluated analytically at $T \to \infty$,
e.g. $m_0^s=\langle jj\rangle/L$, $m_2^s=\langle [H,j][H,j]\rangle/L$
etc.  One gets $m_0^s=J^2/8$ and
\begin{eqnarray}
m_2^s &=& \frac{J^2}{16}[J^2\Delta^2 + 4 \langle b^2\rangle ],\nonumber\\ 
m_0^\epsilon &=& \frac{J^2}{32}[(1+2\Delta^2)J^2 + 2 \langle b^2\rangle ],
\nonumber\\
m_2^\epsilon &=& \frac{J^4}{64} (3+ 10\Delta^2) \langle b^2\rangle
+\frac{J^2}{16} (\langle b^4\rangle -\langle b^2 \rangle^2),
\label{mom}
\end{eqnarray}
where $\langle b^2 \rangle=W^2/12, \langle b^4 \rangle= W^4/80$.
Eqs.(\ref{mom}) reveal the difference between spin and thermal
transport since the finite dispersion
$\delta^\epsilon=\sqrt{m_2^\epsilon/m_0^\epsilon}$ of $\kappa(\omega)$
is induced solely by $W>0$ whereas $\delta^s=\sqrt{m_2^s/m_0^s}$
remains finite even for $W=0$.

Although the lowest frequency moments can serve as a reference, they
are insufficient to reveal the most challenging $\omega \to 0$
behavior. For the latter we have to rely on numerical
calculations. The most favorable case for simulations on a finite size
lattice is the strong disorder limit where we expect the localization
length $\xi$ to be shortest.  In the following we consider $W=2$,
where an estimate of $T=0$ localization length $\xi$ exists in the
literature\cite{eckern}, which suggests that $\xi$ is less than 10
sites in the cases we are studying.

In Fig.~1 and Fig.~2 we present results for $\sigma(\omega)$ and
$\kappa(\omega)$, respectively.  The data for $L=14$, with a Hilbert
space dimension of 3432 states in the $S^z=0$ subsector, were obtained
by exact (full) diagonalization. The $\delta$-peaks at the excitation
frequencies are binned in windows $\delta\omega=0.01$, which also
gives the frequency resolution of the spectra.  There is an average
over $N_r=10$ random field configurations.  For $L=16 - 24$ the
MCLM\cite{mclm} is used, particularly suitable for high $T (\gg J)$,
with typically $2000$ Lanczos steps and random-configuration sampling
$N_r>100$. In the same plots, we show the noninteracting case for
$L=1000$ and averaging $N_r=1000$, where we expect the
dc conductivities $\sigma_{dc}=\sigma(\omega\rightarrow 0)$ and
$\kappa_{dc}=\kappa(\omega\rightarrow 0)$ to vanish. All the spectra
are normalized to a unit integral.

\begin{figure}[htb]
\includegraphics[angle=0, width=.75\linewidth]{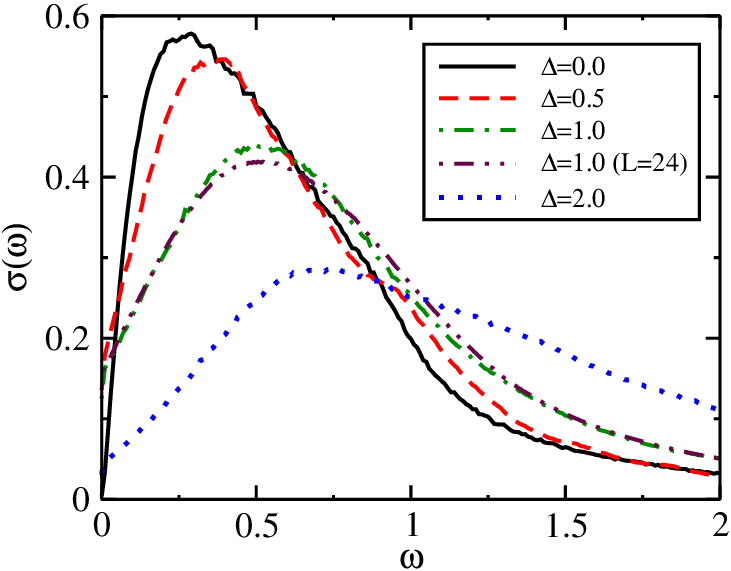}
\caption{Dynamical spin conductivity $\sigma(\omega)$ at $T\to \infty$
  for local disorder $W=2$ and various $\Delta$ (curves normalized to
  unity), evaluated via ED ($L=14$) and for $\Delta=1$ also via MCLM
  ($L=24$).}
\label{fig1}
\end{figure}

\begin{figure}[htb]
\includegraphics[angle=0, width=.75\linewidth]{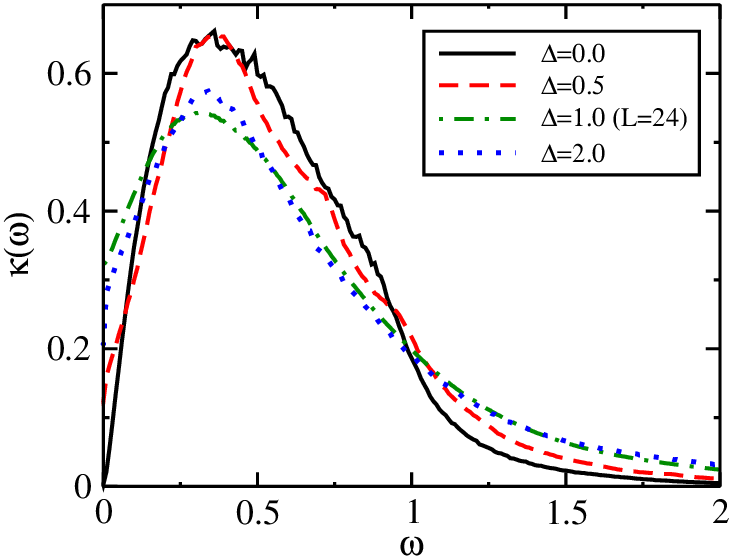}
\caption{Dynamical thermal conductivity $\kappa(\omega)$ at $T\to
\infty$ for $W=2$ and various $\Delta$ (curves normalized to unity), 
evaluated via ED ($L=14$) and MCLM ($L=24$)}.
\label{fig2}
\end{figure}

Data for various sizes $L$ indicate a convergence of the finite-size
effects at moderate sizes $L>16$ (at least for $\Delta>0.5$) for the
chosen rather strong disorder $W=2$. In particular, we show in Fig.~1
comparison of $L=14$ and $L=24$ results for $\Delta=1$. In Fig.~2, for
the same parameters, the curves are also nearly indistinguishable for
$L$ between $14$ and $24$ sites.

These results clearly reveal that apart from the XY limit, $\Delta=0$,
the system is conducting, i.e. both spin $\sigma_{dc}$ and thermal
$\kappa_{dc}$ d.c. conductivities are finite. Nevertheless, due to the
large disorder $W$ the dynamics is non Drude-like, since the maximum
of $\sigma(\omega)$ as well as of $\kappa(\omega)$ appears at a finite
$\omega^*>0$, in analogy with the localization at $\Delta=0$. Hence,
at $\Delta>0$ and large $W$ we are dealing with pseudo-localized
dynamics\cite{huse2,zpp}.  Another novel feature of this regime
appears to be a generic (nonanalytic) cusp-like behavior at low
frequencies, $\sigma(\omega)\simeq \sigma_{dc}+\alpha |\omega|$,
$\kappa(\omega)\simeq \kappa_{dc}+\gamma |\omega|$, for which so far
we cannot offer an analysis. It might be attributed to long-time tail 
effects although, in such a case, 
the low frequency drop of the conductivity 
was found to be only a few percent\cite{wilke} and not by an order 
of magnitude  as in our case.
Such a frequency dependence is strongly reminiscent of the behavior in 
strongly disordered 2D system as has been analyzed theoretically and 
observed experimentally\cite{gold}.

Apart from a qualitative similarity between $\sigma(\omega)$ and
$\kappa(\omega)$ in Figs.~1,2 there are also some differences.
$\sigma(\omega)$ is more sensitive to $\Delta$, as it is already
evident from the moments, Eqs.(\ref{mom}), and corresponding
$\delta^s$.  They originate from the fact that even at $W=0$
$\sigma(\omega \to 0)$ changes qualitatively at $\Delta=1$, not being
the case for $\kappa(\omega \to 0)$\cite{znp}.

As our numerical simulations indicate, a similar qualitative behavior
persists by decreasing the disorder to $W=1.0$ (not shown).  With
decreasing $W$ the pseudo-localized form gives way to a more
Drude-like form with $\omega^* \to 0$ and strongly increased
$\sigma_{dc},\kappa_{dc}$. However, reducing the disorder further we
are running to long localization lengths (for $\Delta=0$) and in
general less controllable finite size effects preventing reliable
conclusions.

The next issue is the temperature dependence of the dynamical (in
particular dc) conductivity and the eventual existence of a critical
temperature $T_c$ below which the system becomes
insulating\cite{mirlin,alt}. To study this question we employed the ED
method for $L=14$ (using $N_r=10$) and the FTLM for $L=16-20$ with
$200-400$ Lanczos steps for high frequency resolution and $N_r \sim
100$. The results are qualitatively similar whereby the FTLM, properly
interpolating between the $T=0$ (ground state) Lanczos method and
$T>0$ behavior, is more reliable for small $T<0.5$ due to larger $L$
and more dense low energy spectra.  Results for $\sigma(\omega)$ and
$\kappa(\omega)$ in the isotropic case $\Delta=1$ and at fixed $W=2$
are shown in Figs.~3,~4 for various $T=0 - 2$ and $L=20$ (being
essentially equal to the results obtained for $L=16$). The data again
indicate that $\sigma_{dc}$ and $\kappa_{dc}$ remain finite at all
$T>0$ vanishing only at $T=0$. A rather abrupt drop of $\sigma_{dc}$
appears at $T\sim 0.1$ which is however in the range of finite-size
temperature $T_{fs}$ (for available $L=20$) below which the FTLM
results are not to be trusted\cite{ftlm}. These data suggest a zero 
critical temperature of localization-delocalization transition, although 
of course we cannot exclude an exponentially small one, which is beyond 
the reach of actual numerical simulations.

\begin{figure}[htb]
\includegraphics[angle=0, width=.75\linewidth]{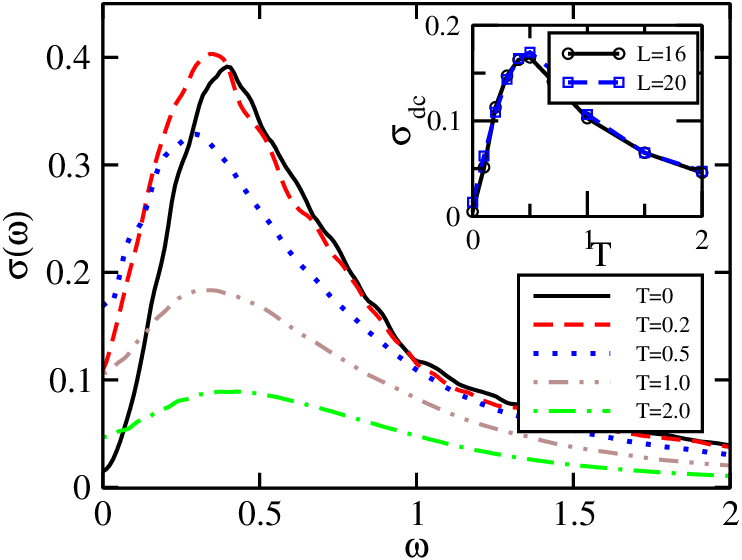}
\caption{Spin 
conductivity $\sigma(\omega)$ for $\Delta=1$ and $W=2$ for various $T$.}
\label{fig3}
\end{figure}

\begin{figure}[htb]
\includegraphics[angle=0, width=.75\linewidth]{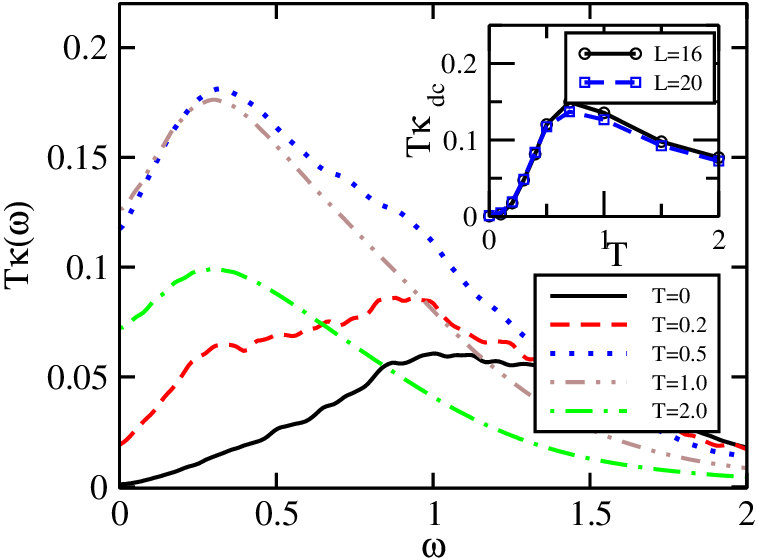}
\caption{Thermal
conductivity $\kappa(\omega)$ for $\Delta=1$ and $W=2$ for various $T$.}
\label{fig4}
\end{figure}

In connection with existing 1D magnetic compounds\cite{hess} more
relevant appears to be the spin-$1/2$ (anisotropic) Hamiltonian with 
bond disorder, i.e. disorder in exchange couplings, 
\begin{equation}
H=\sum_l J_{l,l+1} (S^x_lS^x_{l+1}+S^y_lS^y_{l+1}+\Delta
S^z_lS^z_{l+1}),
\label{bham}
\end{equation}
where $J_{l,l+1}=J(1-s_{l,l+1})$ and we assume $-W/2 < s_{l,l+1} <
W/2$ uniformly distributed random numbers. 
Such disorder can be induced, e.g., by coupling to static 
lattice displacements\cite{lou}.
The local spin current is
now $j_{l,l+1}= J_{l,l+1} (S^x_lS^y_{l+1}-S^y_l S^x_{l+1})$ while the
energy current is given by
\begin{equation}
j^{\epsilon}=\sum_l J_{l-1,l}J_{l,l+1} j^{\epsilon}_l.
\label{bje}
\end{equation}
An open question is whether a 1D spin chain with bond
disorder\cite{huse1} behaves qualitatively different to the site
disorder discussed above.  Our results indicate that it is not the
case.

\begin{figure}[htb]
\includegraphics[angle=0, width=.75\linewidth]{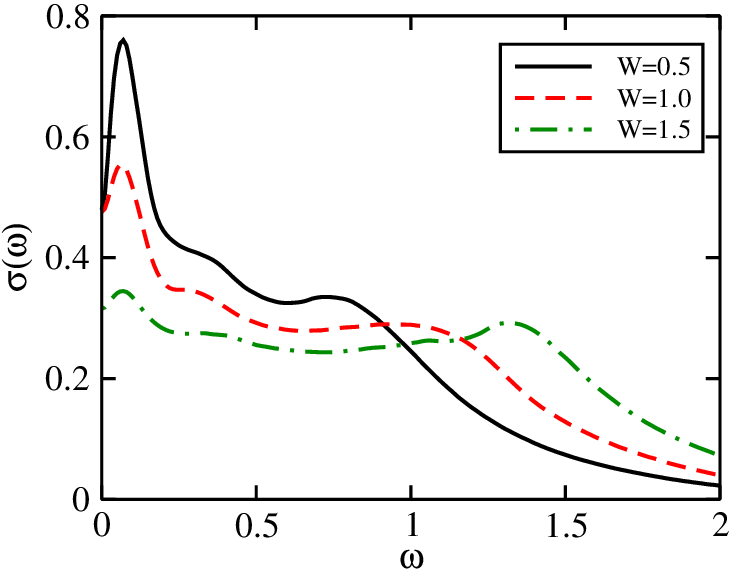}
\caption{$T \to \infty$ results for $\sigma(\omega)$ and for
  $\Delta=1$ and different disorder $W$ (curves are normalized).}
\label{fig5}
\end{figure}

\begin{figure}[htb]
\includegraphics[angle=0, width=.75\linewidth]{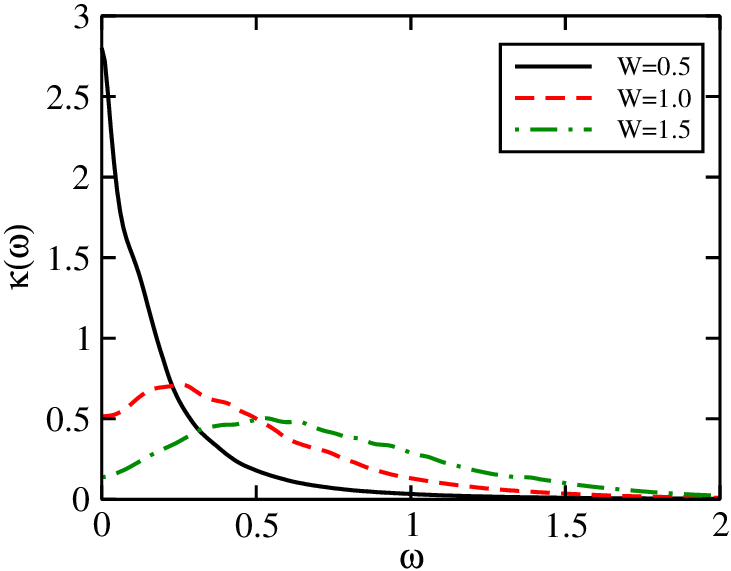}
\caption{$T \to \infty$ results $\kappa(\omega)$ for $\Delta=1$ and
different $W$ (curves are normalized).}
\label{fig6}
\end{figure}

In Figs.~5,~6 we present $T \to \infty$ results for $\sigma(\omega)$ as
well as $\kappa(\omega)$ for $\Delta=1$ and different bond disorder
strengths $W=0.5, 1, 1.5$. Results were obtained using the MCLM method
on $L=20$ sites. Results for larger $W=1, 1.5$ are well converged with
size and clearly indicate that we are again dealing with finite
dc limits $\sigma_{dc}>0$ and $\kappa_{dc}>0$. With respect to the
site disorder case in Figs.~1,~2 there are similarities but also
differences: a) for bond disorder we are restricted to $W<2$ to have a
meaningful model without a possibility of a broken bond, b) the
pseudo-localization is less pronounced at least for $\sigma(\omega)$
and shows up only closer to $W=2$, e.g. for $\kappa(\omega)$ at
$W=1.5$, c) $\kappa(\omega)$ in Fig.~6 reveals a quite abrupt
crossover with disorder strength, from a Drude-like response (at
$W=0.5$) to a localized-like one with $\omega^*>0$ at $W=1.0$, d) at
least for $\sigma(\omega)$ two energy scales are evident in Fig.~5 
which are not
present in the random-field case.

In conclusion, our results of numerical simulations on the interplay of
disorder and correlations in the spin and thermal transport within
Heisenberg spin chains can be summarized by the following scenario: (a)
finite random-field disorder $W>0$ induces localization and vanishing
dc transport at any $T$ in the XY limit, corresponding to
noninteracting fermions\cite{g4}, and as well generally at
$T=0$\cite{eckern,doty} (for $\Delta>0$ considered here); (b) apart
from the latter two limits the system appears to behave as a normal
conductor with finite $\sigma_{dc}>0,\kappa_{dc}>0$ both for various
$\Delta>0$ and $T>0$; in particular, we do not find any evidence for
a phase transition by varying $T$ or $W$; (c) dynamical transport (at least
for larger disorder) reveals a generic cusp-like nonanalytic behavior
for $\omega$, analogous to long-time tails in classical dynamical
systems in low-dimensional \cite{tail} or 2D strongly disordered systems; 
(d) with increasing disorder the system
reveals a crossover from the Drude-like to a pseudo-localized dynamics
with very low dc $\sigma_{dc},\kappa_{dc}$\cite{zpp}; and (e) similar
conclusions seem to hold for the bond disorder.

Clearly, several caveats are in order. The considered cases mostly
correspond to substantial disorder, where the finite size effects are
well under control and results converged within available $L$, at
least for $T>T_{fs}$ and not too small $\Delta >0$. Also, numerical
results cannot exclude the localization on a very long scale $\xi \gg
L$ although we do not find any signature of such a development.

This work was supported by the FP6-032980-2 NOVMAG project and 
by the Slovenian Agency grant No. P1-0044.

\end{document}